\documentclass{article}
\newcommand{\bfnabla}{\mbox{\boldmath $\nabla$}}
\newcommand{\mod}{\mbox{mod }}
\renewcommand{\Re}{\mbox{\rm Re}}
\renewcommand{\Im}{\mbox{\rm Im}}
\begin{document}
\title{Derivation of the Symmetry Postulates for Identical Particles from Pilot-Wave Theories}
\author{Guido Bacciagaluppi\footnote{Department of Philosophy, University of California at Berkeley, 
314 Moses Hall, Berkeley, CA 94720-2390, U.S.A. (e-mail: galuppi@socrates.berkeley.edu).}} 
\date{2 February 2003}
\maketitle
\begin{abstract}
The symmetries of the wavefunction for identical particles, including anyons, are given a rigorous
non-relativistic derivation within pilot-wave formulations of quantum mechanics. In particular, parastatistics are 
excluded. The result has a rigorous generalisation to $n$ particles and to spinorial wavefunctions. The relation to 
other non-relativistic approaches is briefly discussed.  
\end{abstract}

\section{Quantum indistinguishability}\label{quantum}
It is well known that identical particles in standard (non-relativistic) quantum mechanics are 
characterised by symmetry conditions on the wavefunction. For spinless wavefunctions, these are
given by
  \begin{equation}
    \psi({\bf x}_1,\ldots,{\bf x}_i,\ldots,{\bf x}_j,\ldots,{\bf x}_n,t)=e^{i\gamma}\psi({\bf x}_1,\ldots,{\bf x}_j,\ldots,{\bf x}_i,\ldots,{\bf x}_n,t),
    \label{1}
  \end{equation}
where $\gamma=0\,(\mod{2\pi})$ for bosons (symmetry), 
and $\gamma=\pi\,(\mod{2\pi})$ for fermions (antisymmetry). In two dimensions,
it is possible for $\gamma$ to be arbitrary (anyons), where the value depends on the homotopy 
class of the path along which the particles are exchanged, and the wavefunction is multi-valued.
For spinorial wavefunctions (which we discuss only in three dimensions) one requires total 
symmetry or antisymmetry under simultaneous exchange of both spatial and spinorial indices 
(bosons or fermions).\footnote{The possibility of anyons was first pointed out by 
Leinaas and Myrheim (1977). They were described in detail by Goldin {\em et al.} (1981)
and by Wilczek (1982),
and appear to be relevant to the explanation of the fractional quantum Hall effect (Arovas 
{\em et al.}, 1982, Laughlin, 1983, Tsui {\em et al.}, 1982).
}

The derivation of these symmetry conditions from
first principles, however, is not self-evident. As stressed in the classic discussion by 
Messiah and Greenberg (1964), 
the empirical content of quantum indistinguishability is captured by the requirement that 
all observables commute with the permutation operators,
  \begin{equation}
    [A,\Pi_{ij}]=0,\quad\forall A,i,j.
    \label{2}
  \end{equation}
But this merely leads to the conclusion that wavefunctions describing 
indistinguishable particles should lie in subspaces of Hilbert space
that are invariant under all elements of an irreducible representation 
of the permutation group. Since for $n>2$ particles the permutation group 
is non-commutative, one obtains in addition to the usual one-dimensional 
representations also higher-dimensional ones, connected with the so-called
{\em parastatistics}, and which have never been observed experimentally.

It is the purpose of this paper to give a natural proof that parastatistics are excluded 
and a derivation of the three known symmetry types in the framework of two pilot-wave 
theories, namely de Broglie--Bohm theory (de Broglie, 1928, Bohm, 1952),
and Nelson's (1966) stochastic mechanics. 
In non-relativistic quantum theory, this is one of the most rigorous proofs to date, in
particular in int extension to spinors. The two pilot-wave theories are sketched in 
Section \ref{bohm}. The main result follows in Section \ref{main} for two spinless particles,
and is generalised to $n$ particles and to spinors in Sections \ref{n} and \ref{spin}, respectively.  
Finally, in Section \ref{topological}, we briefly examine the relation of our 
approach to the topological approaches of Laidlaw and C. DeWitt (1971)
and Leinaas and Myrheim (1977),
and to the approach of Goldin {\em et al.} (1980, 1981).

An independent treatment of this question by D\"{u}rr {\em et al.\ }(in preparation) 
is forthcoming, which explicitly includes spinorial wavefunctions, and gives a label-free 
description of identical particles in the Bohm framework. For other partial or related results 
in de Broglie--Bohm theory see Schneider (1995),\footnote{Schneider shows under that under 
quite general conditions, two wavefunctions satisfying the same Schr\"{o}dinger equation and 
generating the same de Broglie velocity field are proportional to each other. Applying this 
to identical particles, he obtains a result similar to ours (for scalar wavefunctions).} 
Sj\"{o}qvist and Carlsen (1995) and Brown {\em et al.\ }(1999),
and in stochastic mechanics see Nelson (1985) and Goldstein (1987).

\section{Pilot-wave theories}\label{bohm}
The development of wave mechanics, with both de Broglie and Schr\"{o}dinger, was based on the 
optico-mechanical analogy. For de Broglie (1928),
the Hamilton--Jacobi action, which is related to the canonical momentum of particle $i$ via
  \begin{equation}
    {\bf p}_i = \bfnabla_i S,
    \label{3}
  \end{equation}  
should be interpreted as the geometric limiting case of the phase of a wave, for which a wave equation (on configuration space) had meanwhile been derived by Schr\"{o}dinger.
While in Schr\"{o}dinger's formulation of wave mechanics, as in wave optics, one does away with light rays or particle trajectories, de Broglie's formulation 
is a dynamical theory of point masses, a pilot-wave theory, defined by 
a {\em guidance equation} of the form (\ref{3}) and the Schr\"{o}dinger equation for $\psi=Re^{iS/\hbar}$. In the case of non-vanishing vector 
potential, the guidance equation reads 
  \begin{equation}
    m_i\dot{{\bf x}}_i - q_i{\bf A}({\bf x}_i)= \bfnabla_i S ,
    \label{4}
  \end{equation}
or
  \begin{equation}
    \dot{\bf x}_i  = \frac{1}{m_i}(\bfnabla_i S + q_i{\bf A}({\bf x}_i)),
    \label{4a}
  \end{equation}
and is thus gauge invariant. 

From the Schr\"{o}dinger equation one standardly derives the two coupled equations for $R^2$ and $S$:
  \begin{eqnarray}
    \partial_t R^2  & = & -\sum_i\bfnabla_i\cdot (R^2\frac{1}{m_i}\bfnabla_i S),
    \label{5}        \\
    \partial_t S  & = & -\sum_i\frac{(\bfnabla_i S)^2}{2m_i} -V+\sum_i\frac{\hbar^2}{2m_i}\frac{\Delta_i R}{R},
    \label{6}
  \end{eqnarray}
and accordingly with vector potential. Equation (\ref{5}) is a continuity equation for $R^2$, while equation (\ref{6}) has the form of a  Hamilton--Jacobi equation with 
an additional term, sometimes called the quantum potential,
  \begin{equation}
    Q=-\sum_i\frac{\hbar^2}{2m_i}\frac{\Delta_i R}{R}.
    \label{7}
  \end{equation}
This explains intuitively results such as electron diffraction and interference, as can be seen qualitatively from the shape 
of the quantum potential (see e.g.\ Bohm and Hiley, 1993, p. 34). Notice that the quantum potential factorises if and only if the wavefunction does.

After a lull of twenty-five years, the theory was rediscovered by Bohm (1952)
and revitalised through his discovery that it allows a consistent 
quantum description of measurement and of the phenomenology of collapse. We now sketch the latter description, which we need for Section 
\ref{spin}. 

During a measurement interaction, eigenstates $\psi_i(x)$ of some self-adjoint operator of a system (call it the particle) become 
coupled with essentially non-overlapping states $\Phi_i(y)$ of some other system (call it the apparatus): 
  \begin{equation}
    \sum_i \lambda_i \psi_i(x)\Phi_i(y).
    \label{8}
  \end{equation}
This means that during the measurement interaction, the configuration space of the composite system is carved up into regions separated by
high quantum potential barriers, corresponding to the different components of (\ref{8}). If the components are not reinterfered, the 
future trajectory of the particle will be guided, via (\ref{4a}), only by one of the states $\psi_i(x)$, the {\em effective wavefunction} of the 
particle, as if the wave had physically collapsed. Given a fixed experimental arrangement, it is the initial configuration of the system that determines
which component is the effective wavefunction. In this sense, one can talk of the theory as a hidden variables theory. In general, however, 
the result depends also on the experimental arrangement, so that the standard arguments against such theories do not apply (Bell, 1987, 
esp.\ Chaps 1 and 18). 

Application of this model to position and momentum measurements already yields a qualitative understanding of
uncertainty, and application to parallel Stern--Gerlach measurements on particles in the singlet state yields strict 
anticorrelations (see Section \ref{spin} for the relevant generalisation of the theory to spinors). The quantitative aspects of quantum 
mechanics are obtained by assuming that particle positions in an ensemble are distributed according to the quantum distribution
$R^2=|\psi|^2$, which will then be true at all times (under a unitary evolution the particle distribution satisfies the same continuity 
equation as $R^2$, and in a measurement the preceding analysis shows why no other subensembles can be selected other than
those characterised by one of the distributions $|\psi_i|^2$). This distribution plays a role analogous to that of equilibrium distributions in classical 
statistical mechanics. Indeed, the analogy between the two theories stretches from justifications of equilibrium down to cousins of 
Maxwell's demon.

Standard references to de Broglie--Bohm theory, in addition to the original papers, are the textbooks by Bohm and Hiley (1993),
Holland (1993),
D\"{u}rr (2001)
and the forthcoming one by Valentini (in preparation).
A concise non-technical overview is given by Goldstein (2001).
The model of spin to be used in Section \ref{spin} is due to Bell 
(1987, Chap.\ 19), which contains several other papers providing insights into, extending and generally promoting the approach.

Nelson's (1966, 1985) stochastic mechanics 
can also be seen as a pilot-wave theory, although this is not his own understanding. One modifies 
de Broglie--Bohm theory by assuming that the particle undergoes not a deterministic evolution but a {\em diffusion process},
in which the guidance equation is given by the following It\^{o} equation:
  \begin{equation}
    d{\bf x}_i = {\bf b}_i dt+d\mbox{\boldmath $\omega$}_i,   
    \label{9}
  \end{equation}
with
  \begin{equation}
    {\bf b}_i=\frac{1}{m_i}\bfnabla_i S+\frac{q_i}{m_i}{\bf A}
    +\frac{\hbar}{2m_i}\frac{\bfnabla_i R^2}{R^2}, 
    \label{10}
  \end{equation}
and where $d\mbox{\boldmath $\omega$}_i$ is a Wiener process with
  \begin{equation}
    \overline{d\mbox{\boldmath $\omega$}_i}=0,\quad 
    \overline{(d\mbox{\boldmath $\omega$}_i)^2}=\frac{\hbar}{m_i}dt.
    \label{11}
  \end{equation}
Here, ${\bf b}_i$ is the drift velocity of particle $i$, and $\nu_i:=\frac{\hbar}{2m_i}$ its diffusion coefficent. 

By inserting the distribution $R^2$, the corresponding Fokker--Planck equation reduces to the continuity equation
(\ref{5}), so that one sees both that $R^2$ is the equilibrium distribution also for Nelson's theory, and that the average
velocity of the particles at equilibrium is their de Broglie--Bohm velocity (\ref{4a}). Thus, one can consider Nelson's theory as 
de Broglie--Bohm theory with added noise (see also Bohm and Hiley, 1993,
Sects 9.5 ff.).

\section{Case of two spinless particles}\label{main}
Since pilot-wave theories include more structure than the standard formulation of quantum mechanics, namely particle trajectories, 
it is now possible to formulate stronger conditions of indistinguishability than merely requiring (\ref{2}), and this will allow us to derive
all the standard symmetry properties for wavefunctions. We start with the case of two spinless particles. Since any Hamiltonian will 
be symmetrical, it suffices to establish the result at a single instant in time; it will then hold for all instants. We treat the phase $S$ and 
amplitude $R$ in succession. We shall allow $S$ in general to be multi-valued, while $R$ is obviously single-valued and positive. Notice, 
however, that although $S$ may be multi-valued, $\partial_t S$ and all the $\bfnabla_i S$ need to be uniquely defined, in order for $Re^{iS/\hbar}$
to solve the Schr\"{o}dinger equation, in particular (\ref{5}) and (\ref{6}). (A unique $\bfnabla_i S$ is also needed for the corresponding guidance equation to be well-defined.) In the following, we shall set $m_i=m=1$ and $q_i=q=1$ for all $i$.

The treatment of $S$ is identical in de Broglie--Bohm theory and in Nelson's mechanics. We impose a symmetry condition on the velocities
of the particles (respectively, on the average velocities), as given by (\ref{4a}):
  \begin{equation}
    \bfnabla_1 S(\xi_1,\xi_2,t)+{\bf A}(\xi_1,t)\Big|_{\xi_1={\bf x},\xi_2={\bf y}} = \bfnabla_2 S(\xi_1,\xi_2,t)+{\bf A}(\xi_2,t)\Big|_{\xi_1={\bf y},\xi_2={\bf x}}.
    \label{12}
  \end{equation}
That is, the velocity (average velocity) of particle 1 in a given configuration is equal to that of particle 2 in the configuration with the particles exchanged. 
This is the natural requirement of {\em indistinguishability} at the level of particle trajectories.
By renaming the variables on the right-hand side, we see that the vector potential drops out, and we obtain:
  \begin{equation}
    \bfnabla_1 \Big(S(\xi_1,\xi_2,t) - S(\xi_2,\xi_1,t)\Big)\Big|_{\xi_1={\bf x},\xi_2={\bf y}}=0.
    \label{13}
  \end{equation}
Similarly we obtain:
  \begin{equation}
    \bfnabla_2 \Big(S(\xi_1,\xi_2,t) - S(\xi_2,\xi_1,t)\Big)\Big|_{\xi_1={\bf x},\xi_2={\bf y}}=0.
    \label{14}
  \end{equation}
In other words, we have (with $2D$-dimensional notation for the $\bfnabla$ operator, $D$ the dimension of space):
  \begin{equation}
    \bfnabla \Big(S({\bf x},{\bf y},t) - S({\bf y},{\bf x},t)\Big) =0.
    \label{15}
  \end{equation}

Were $S$ everywhere defined, we would conclude that $S({\bf x},{\bf y},t) = S({\bf y},{\bf x},t) + \gamma\, (\mod 2\pi)$, and in fact $\gamma=0$ for all $t$
(setting ${\bf x}={\bf y}$). However, $S$ is undefined where $R$ has a zero. For example, if the two particles are trapped in two disjoint boxes with infinite 
potential walls, the region $\{R\neq 0\}$ is disconnected, and $\gamma$ can be arbitrary. Indeed, if $\psi$ and $\varphi$ are single-particle wavefunctions 
with support in the two boxes, respectively, such that the symmetric combination $\psi(\xi_1)\varphi(\xi_2)+\varphi(\xi_1)\psi(\xi_2)$ solves the   
Schr\"{o}dinger equation, the two terms have disjoint support in the configuration space, and solve the Schr\"{o}dinger equation separately. Thus any
wavefunction
  \begin{equation}
    \psi(\xi_1)\varphi(\xi_2)+\alpha e^{i\beta}\varphi(\xi_1)\psi(\xi_2)
    \label{16}
  \end{equation}
is also a solution. (Notice that not only the phase difference may be arbitrary, but also the symmetry of the amplitude may be destroyed. Incidentally, (\ref{16}) furnishes
an example in which there is more than one equilibrium measure for the particle distribution.) 

We must thus require, as is implicit also in the standard 
requirement (\ref{2}), that the particles be indistinguishable under {\em all possible circumstances}, in particular if we open the boxes. (This and similar ways of speaking in the following
do not imply reference to actual laboratory procedures, but to the fact that (\ref{15}) needs to hold at all times even with arbitrary time-dependent symmetric potentials, magnetic fields 
etc.) 

We still may not assume, however, that the spreading of the wave function makes $R$ everywhere different from zero, because we know that the diagonal set $\{{\bf x}={\bf y}\}$ 
may be exceptional, from the example of antisymmetric wavefunctions. (It is also exceptional from the point of view of the pilot-wave dynamics, as discussed below in Section 
\ref{topological}.) We omit the case $D=1$, for which the set $\{{\bf x}\neq{\bf y}\}$ is disconnected. For $D\geq 2$ instead this set is connected. Given the freedom to vary $V$ etc.,
we shall thus assume that there
is an open time interval in which $R$ is different from zero on the set $\{{\bf x}\neq{\bf y}\}$ (or at least a connected subset). On this interval we can indeed
conclude
  \begin{equation}
    S({\bf x},{\bf y},t) = S({\bf y},{\bf x},t) + \gamma\, (\mod 2\pi).
    \label{17}
  \end{equation}

For $D\geq 3$ the set $\{{\bf x}\neq{\bf y}\}$ is simply connected (or we may assume that so is the connected subset). Thus $S$ is single-valued ($\mod 2\pi$). Interchanging
${\bf x}$ and ${\bf y}$ in (\ref{17}), we see that $\gamma$ equals $0$ or $\pi$ (bosons or fermions). When $D=2$ the set $\{{\bf x}\neq{\bf y}\}$ is multiply connected. The (equal-time) 
paths $\lambda$ from $({\bf x},{\bf y})$ to $({\bf y},{\bf x})$ fall into different homotopy classes, depending on whether the two particles pass each other on the left or on the right (call 
these `simple exchange paths'), or on whether and how often they wind around each other in either sense. Any exchange path is homotopic to a concatenation of simple exchange paths. 
Thus for $D=2$ the phase difference along an exchange path is independent of ${\bf x}$ and ${\bf y}$, and is equal to a whole number $n$, depending on the homotopy class, times an 
arbitrary constant $\gamma$ (anyonic phase). This holds over a whole time interval, and the question arises whether $\gamma$ may be time-dependent. Were this the case, 
however, $\partial_t S$ would be ill-defined. Thus $\gamma$ is independent of $t$, for all $D\geq 2$. 

We now establish that $R$ is symmetric. The second proof given below covers both de Broglie and Bohm's theory and Nelson's, but we note that there is
a more direct proof in Nelson's theory. (Both proofs hold for all $D\geq 2$.) 
Indeed, we require not only that the average velocity (\ref{4a}), but also the drift velocity (\ref{10}) be symmetric. The 
two conditions combined yield: 
  \begin{equation}
    \frac{\bfnabla R^2({\bf x},{\bf y},t)}{R^2({\bf x},{\bf y},t)} = \frac{\bfnabla R^2({\bf y},{\bf x},t)}{R^2({\bf y},{\bf x},t)}.
    \label{19}
  \end{equation}
We take any instant from the interval for which $R\neq 0$ on $\{{\bf x}\neq{\bf y}\}$ (or the connected subset), and thus again consider (\ref{19}) on this set ($R$ is already symmetric on 
the complement). From (\ref{19}) we have:  
  \begin{equation}
    \bfnabla\log R^2({\bf x},{\bf y},t) = \bfnabla\log R^2({\bf y},{\bf x},t).
    \label{20}
  \end{equation}
Thus,
  \begin{equation}
    \bfnabla\log R({\bf x},{\bf y},t) = \bfnabla\log R({\bf y},{\bf x},t),
    \label{21}
  \end{equation}
and since for $D\geq 2$ the region is connected,
  \begin{equation}
    \log R({\bf x},{\bf y},t) = \log R({\bf y},{\bf x},t)+\delta,
    \label{22}
  \end{equation}
and
  \begin{equation}
    R({\bf x},{\bf y},t) = e^{\delta}R({\bf y},{\bf x},t).
    \label{23}
  \end{equation}
It follows that
  \begin{equation}
    R({\bf x},{\bf y},t) = e^{2\delta}R({\bf x},{\bf y},t),
    \label{24}
  \end{equation}
and since $R$ is single-valued by definition, $e^{2\delta}=1$, or
  \begin{equation}
    R({\bf x},{\bf y},t) = \pm R({\bf y},{\bf x},t);
    \label{25}
  \end{equation}
and since it is positive,
  \begin{equation}
    R({\bf x},{\bf y},t) = R({\bf y},{\bf x},t),
    \label{26}
  \end{equation}
for any instant of the time interval under consideration (both on $\{{\bf x}\neq{\bf y}\}$ and on $\{{\bf x}={\bf y}\}$). Thus, in both phase and amplitude the wavefunction has the
desired symmetries on a certain time interval, and since $H$ is symmetric, it has them at all times.

For the case of de Broglie and Bohm's theory, we do not have the additional condition (\ref{19}), but we use the fact that on the given time interval $S$ is symmetric (up to 
a time-independent $\gamma$), and $R$ and $S$ are coupled via (\ref{5}) and (\ref{6}), with symmetric $V$. Since we can vary the Hamiltonian, we choose the case of zero 
magnetic field, so in an appropriate gauge ${\bf A}=0$. (This is not only for ease of calculation, but also because at one point in Section \ref{spin} we shall need precisely 
this case.) We introduce the notation $\tilde{S}({\bf x},{\bf y},t) := S({\bf y},{\bf x},t)$ and $\tilde{R}({\bf x},{\bf y},t) := R({\bf y},{\bf x},t)$.

First of all, since $Re^{iS/\hbar}$ and $\tilde{R}e^{i\tilde{S}/\hbar}$ both solve the Schr\"{o}dinger equation and thus (\ref{6}), and since all of $\partial_t S$, $\bfnabla S$ and $V$ 
are symmetric on the given time interval, it follows that the quantum potential (\ref{7}) is symmetric:
  \begin{equation}
    -\frac{\hbar^2}{2}\frac{\Delta R}{R} = -\frac{\hbar^2}{2}\frac{\Delta \tilde{R}}{\tilde{R}},
    \label{27}
  \end{equation}
or
  \begin{equation}
    \tilde{R} \Delta R- R\Delta\tilde{R} = 0
    \label{28}
  \end{equation}
identically, and thus also
  \begin{equation}
    \partial_t\left(\tilde{R} \Delta R- R\Delta\tilde{R}\right) = 0.
    \label{29}
  \end{equation}

From the continuity equation (\ref{5}), on the region $R\neq 0$, one has
  \begin{equation}
    \partial_t R = -\bfnabla R \bfnabla S - \frac{R}{2}\Delta S,
    \label{30}
  \end{equation}
and similarly for $\tilde{R}$ and $\tilde{S}$, since they also satisfy the continuity equation. Substituting into (\ref{29}) one has:
  \begin{equation}
    \begin{array}{rcl}
      \partial_t \Big( \tilde{R} \Delta R- R\Delta\tilde{R} \Big)  
      & = &       
      (\partial_t\tilde{R})\Delta R+\tilde{R}\Delta(\partial_t R)
      -(\partial_t R)\Delta \tilde{R}- R\Delta(\partial_t\tilde{R})     
      \\[1ex]                                                                          
      & = &
      \bfnabla R \bfnabla S \Delta \tilde{R} 
      + {\displaystyle \frac{R}{2}}\Delta S \Delta\tilde{R} +   
      R\bfnabla(\Delta\tilde{R})\bfnabla\tilde{S} +
      \\[0.5ex]
      & &
      {\displaystyle 2R(\sum_{m,n}\partial_m\partial_n\tilde{R}\partial_m\partial_n\tilde{S})
      +R\bfnabla\tilde{R}\bfnabla(\Delta\tilde{S}) + } 
      \\[0.5ex]
      & &
      {\displaystyle R\frac{\Delta\tilde{R}}{2}\Delta\tilde{S} +
      R\bfnabla\tilde{R}\bfnabla(\Delta\tilde{S})
      + R\frac{\tilde{R}}{2}\Delta\Delta\tilde{S} -}    
      \\[0.5ex]
      &   &
      \bfnabla \tilde{R} \bfnabla \tilde{S} \Delta R - 
      {\displaystyle \frac{\tilde{R}}{2}}\Delta \tilde{S} \Delta R   -      
      \tilde{R}\bfnabla(\Delta R)\bfnabla S-
      \\[0.5ex]
      & &
      {\displaystyle 2\tilde{R}(\sum_{m,n}\partial_m\partial_n R\partial_m\partial_n S)-
      \tilde{R}\bfnabla R\bfnabla(\Delta S) -}   
      \\[0.5ex]
      & &
      {\displaystyle \tilde{R}\frac{\Delta R}{2}\Delta S -
      \tilde{R}\bfnabla R\bfnabla(\Delta S) -
      \tilde{R}\frac{R}{2}\Delta\Delta S} 
    \end{array}
    \label{31}
  \end{equation}
(where we continue to use $2D$-dimensional notation for $\bfnabla$ and $\Delta$, and $m$ and $n$
also run from $1$ to $2D$).
Simplifying using (\ref{28}) and the symmetry of $\bfnabla S$ and
$\Delta S$ yields
  \begin{equation}
    \begin{array}{lcl}
      \bfnabla S\Big(\bfnabla R \Delta\tilde{R} + R\bfnabla(\Delta\tilde{R})
                -\bfnabla \tilde{R} \Delta R - \tilde{R}\bfnabla(\Delta R)\Big)+
      & &  \\[0.5ex]
      \qquad
      2(\bfnabla\Delta S)(R\bfnabla\tilde{R}-\tilde{R}\bfnabla R) +
      {\displaystyle 2\sum_{m,n}\partial_m\partial_n S
      (R\partial_m\partial_n\tilde{R}-\tilde{r}\partial_m\partial_n R) }  
      & = &   \\[1.5ex]
      (\bfnabla S) \bfnabla(R\Delta\tilde{R}-\tilde{R}\Delta R) + 
      2(\bfnabla\Delta S)(R\bfnabla\tilde{R}-\tilde{R}\bfnabla R)+
      & &  \\[0.5ex]
      \qquad
      {\displaystyle 2\sum_{m,n}\partial_m\partial_n S
      (R\partial_m\partial_n\tilde{R}-\tilde{r}\partial_m\partial_n R) }  
      & = &  0.
    \end{array}
    \label{32}
  \end{equation}
Thus, again because of (\ref{28}),
  \begin{equation}
    (\bfnabla\Delta S)(R\bfnabla\tilde{R}-\tilde{R}\bfnabla R)+
    \sum_{m,n}\partial_m\partial_n S
    (R\partial_m\partial_n\tilde{R}-\tilde{r}\partial_m\partial_n R)=0,
    \label{33}
  \end{equation}
and again taking the time derivative,
  \begin{equation}
    \begin{array}{l}
      \partial_t\Big( (\bfnabla\Delta S)(R\bfnabla\tilde{R}-\tilde{R}\bfnabla
      R)\Big)=    \\[1.5ex]
      \qquad\qquad
      (\bfnabla\Delta\partial_tS) (R\bfnabla\tilde{R}-\tilde{R}\bfnabla R) +
      \bfnabla(\Delta S) \partial_t(R\bfnabla\tilde{R}-\tilde{R}\bfnabla R)+   \\[0.5ex]
      \qquad\qquad
      {\displaystyle \sum_{m,n}(\partial_m\partial_n\partial_t S)
      (R\partial_m\partial_n\tilde{R}-\tilde{r}\partial_m\partial_n R)+}    \\
      \qquad\qquad
      {\displaystyle \sum_{m,n}(\partial_m\partial_n S)
      \partial_t(R\partial_m\partial_n\tilde{R}-\tilde{r}\partial_m\partial_n R)}
      \qquad\qquad = \quad 0.
    \end{array}
    \label{34}
  \end{equation}

In the limit of instantaneously varying $V$, we can affect (via (\ref{6})) the terms containing
$\partial_tS$ without changing the others. Indeed, at any time $t$, we can add an arbitrary 
(symmetric) term $U$ to the potential, thus adding to (\ref{34}) a term of the form
  \begin{equation}
    -(\bfnabla\Delta U) (R\bfnabla\tilde{R}-\tilde{R}\bfnabla R)
    -\sum_{m,n}\partial_m\partial_n U
      (R\partial_m\partial_n\tilde{R}-\tilde{r}\partial_m\partial_n R)
    \label{35}
  \end{equation}
which has to be zero. If we further choose $U$ of the form
  \begin{equation}
    U=u^1(\xi_1^1)+\ldots +u^D(\xi_1^D)+u^1(\xi_2^1)+\ldots +u^D(\xi_2^D),
    \label{36}
  \end{equation}
all mixed derivatives vanish, and the requirement reduces to
  \begin{equation}
    \sum_n\partial_n^3 U (R\partial_n\tilde{R}-\tilde{R}\partial_n R)+
    \sum_n\partial_n^2 U (R\partial_n^2\tilde{R}-\tilde{R}\partial_n^2 R)=0,
    \label{37}
  \end{equation}
where all $\partial_n^2 U$ and $\partial_n^3 U$ can be varied independently. Thus, in particular,
  \begin{equation}
    R \bfnabla\tilde{R}- \tilde{R}\bfnabla R = 0.
    \label{38}
  \end{equation}

From this, on $\{R\neq 0\}$, follows again equation (\ref{21}), and the rest of the proof 
proceeds as in the case of Nelson's mechanics.

\section{Case of $n$ spinless particles}\label{n}
For $n$ particles, one can take over the above arguments largely unmodified, and conclude that for all $i,j$,
  \begin{equation}
    S({\bf x}_1,\ldots,{\bf x}_i,\ldots,{\bf x}_j,\ldots,{\bf x}_n,t)=S({\bf x}_1,\ldots,{\bf x}_j,\ldots,{\bf x}_i,\ldots,{\bf x}_n,t)+\gamma_{\{i,j\}}.
    \label{40}
  \end{equation}
As before, $\gamma_{\{i,j\}}$ is independent of $t$, and also of ${\bf x}_k$ ($k\neq i,j$), because $\partial_tS$ and $\bfnabla S$ would otherwise be ill-defined.
As before, for $D\geq 3$ the configuration space without the set where two or more particles coincide is simply connected, and thus the constant $\gamma_{\{i,j\}}$ 
can take only the two values $0$ or $\pi$, while for $D=2$ it is arbitrary. The only additional step that is required is to show that $\gamma_{\{i,j\}}$ is in fact independent 
of the pair $\{i,j\}$, and, for $D=2$, to discuss the dependence on the homotopy structure of the exchange paths, which is rather more complicated for $n$ particles than it is 
for two.

For $D=3$, the phase difference between $S$ and $\tilde{S}$ is path-independent, and thus $\gamma_{\{i,j\}}$ is uniquely associated with a permutation of the labels $i,j$.
For permutations $\Pi_{ij}$ we have, however:
  \begin{equation}
    \Pi_{ij}=\Pi_{jk}\Pi_{ik}\Pi_{jk},
    \label{41}
  \end{equation}
so that
  \begin{equation}
    \gamma_{\{i,j\}}=\gamma_{\{j,k\}}+\gamma_{\{i,k\}}+\gamma_{\{j,k\}}\, (\mod 2\pi).
    \label{42}
  \end{equation}
Since
  \begin{equation}
    \gamma_{\{j,k\}}+\gamma_{\{j,k\}}=0\, (\mod 2\pi),
    \label{43}
  \end{equation}
we have
  \begin{equation}
    \gamma_{\{i,j\}}=\gamma_{\{i,k\}}\, (\mod 2\pi),
    \label{44}
  \end{equation}
and exchange of one particle with any other yields the same phase difference. It follows that exchange of any two particles yields the same phase difference.

Take $D=2$. For two particles we have said that all exchange paths are homotopic to concatenations of the simple exchange paths along which the two particles do not wind around each other. 
For $n$ particles, two such paths, even if like-handed, are not generally homotopic to each other. The homotopy class further
depends on whether and how many other particles may be enclosed within the path, and on whether and how often any such particles are circled by either of the particles being exchanged.
We can, however, redefine simple exchange paths as those along which the particles being exchanged neither wind around each other nor enclose any other particles. 
One can then easily convince oneself that any exchange path is homotopic to a concatenation of such simple exchange paths. The phase difference along a simple exchange path 
is $\pm\gamma_{\{i,j\}}$, according to whether the path is left-handed or right-handed. What we wish to establish is that this is independent of the pair $\{i,j\}$. And in fact, we can 
mimick the above argument based on permutations. 

We first exchange particles $j$ and $k$ along a simple exchange path, say a left-handed one; then we exchange $i$ and $k$ along another, say also a left-handed one; 
finally we exchange $j$ and $k$ again, this time along a {\em right-handed} simple exchange path. This concatenation is homotopic to a simple 
path exchanging $i$ and $j$ (a left-handed one). In terms of the phase differences along the paths, we obtain:
  \begin{equation}
    \gamma_{\{i,j\}}=\gamma_{\{j,k\}}+\gamma_{\{i,k\}}-\gamma_{\{j,k\}}\, (\mod 2\pi).
    \label{45}
  \end{equation}
That is, even without (\ref{43}), we obtain the result (\ref{44}), now for the phase differences along simple exchange paths (not enclosing any other particles) in two dimensions.

\section{Generalisation for spinorial wavefunctions}\label{spin}
We shall now treat the case of spinorial wavefunctions (in three dimensions). These obey the Pauli equation:
  \begin{equation}
    i\hbar\frac{\partial{\bf \Psi}}{\partial t}=-\sum_i\frac{\hbar^2}{2m_i}(\bfnabla_i-iq_i{\bf A}({\bf x}_i))^2{\bf \Psi}+V{\bf \Psi}+\mu\sum_i{\bf S}_i{\bf B}({\bf x}_i){\bf \Psi},
    \label{45a}
  \end{equation}
where ${\bf S}_i$ is the spin matrix vector acting on particle $i$, and ${\bf B}$ is the (external) magnetic field. Writing spinorial wavefunctions as 
  \begin{equation}
    {\bf \Psi}=\sum_{s_1,\ldots,s_n}\psi_{s_1,\ldots,s_n}|s_1>\ldots |s_n>, 
    \label{47a}
  \end{equation}
the condition to be derived is
  \begin{equation}
    \begin{array}{l}
      \psi_{s_1\ldots s_i\ldots s_j\ldots s_n}({\bf x}_1,\ldots,{\bf x}_i,\ldots,{\bf x}_j,\ldots,{\bf x}_n,t)=           \\[1ex]
      \qquad\qquad\qquad\qquad    e^{i\gamma}\psi_{s_1\ldots s_j\ldots s_i\ldots s_n}({\bf x}_1,\ldots,{\bf x}_j,\ldots,{\bf x}_i,\ldots,{\bf x}_n,t),   
      \label{50}
    \end{array}
  \end{equation}
with $\gamma=0$ for all $i,j$ or $\gamma=\pi$ for all $i,j$ (total symmetry or antisymmetry of the spinorial wavefunction). Notice also that for ${\bf B}=0$, the single 
components of the spinorial wavefunction decouple and all obey the same (symmetric) Schr\"{o}dinger equation (in which ${\bf A}$ can be gauged to zero).

We first generalise the two pilot-wave theories to the case of spinorial wavefunctions, following the most common treatment, due to Bell 
(1987, Chap.\ 19). Notice first of all that de Broglie's guidance equation (\ref{4a}) can be written
  \begin{equation}
    \dot{\bf x}_i  = \frac{\hbar}{m_i}\Im\frac{\psi^*\bfnabla_i\psi}{\psi^*\psi} + \frac{q_i}{m_i}{\bf A}({\bf x}_i ),
    \label{46}
  \end{equation}
and similarly, Nelson's can be written
  \begin{equation}
    d{\bf x}_i = \Big(\frac{\hbar}{m_i}\Im\frac{\psi^*\bfnabla_i\psi}{\psi^*\psi}+\frac{q_i}{m_i}{\bf A}({\bf x}_i )+\frac{\hbar}{m_i}\Re\frac{\psi^*\bfnabla_i\psi}{\psi^*\psi}\Big)dt
   +d\mbox{\boldmath $\omega$}_i.   
    \label{47}
  \end{equation}
The generalisations of (\ref{46}) and (\ref{47}) to spin are given, respectively, by
  \begin{equation}
    \dot{\bf x}_i  = 
    \frac{\hbar}{m_i}\Im\frac{{\displaystyle\sum_{s_1,\ldots,s_n}}\psi_{s_1\ldots s_n}^*\bfnabla_i\psi_{s_1\ldots s_n}}{{\displaystyle\sum_{s_1,\ldots,s_n}}\psi_{s_1\ldots s_n}^*\psi_{s_1\ldots s_n}} 
   + \frac{q_i}{m_i}{\bf A}({\bf x}_i ),
    \label{48}
  \end{equation}
and 
  \begin{eqnarray}
    d{\bf x}_i       & = & 
    \Big(\frac{\hbar}{m_i}\Im\frac{{\displaystyle\sum_{s_1,\ldots,s_n}}\psi^*_{s_1\ldots s_n}\bfnabla_i\psi_{s_1\ldots s_n}}{{\displaystyle\sum_{s_1,\ldots,s_n}}\psi^*_{s_1 \ldots s_n}\psi_{s_1\ldots s_n}}
   +\frac{q_i}{m_i}{\bf A}({\bf x}_i )+                  
                                                                              \nonumber       \\
                             &    &
   \frac{\hbar}{m_i}\Re\frac{{\displaystyle\sum_{s_1,\ldots,s_n}}\psi^*_{s_1\ldots s_n}\bfnabla_i\psi_{s_1\ldots s_n}}{{\displaystyle\sum_{s_1,\ldots,s_n}}\psi^*_{s_1\ldots s_n}\psi_{s_1\ldots s_n}}\Big)dt
   +d\mbox{\boldmath $\omega$}_i,   
    \label{49}
  \end{eqnarray}
which are manifestly independent of the spin basis chosen. (These are the equations needed to analyse the Bell experiments.)

Our methods of Section \ref{main} are {\em not} directly applicable to derive (\ref{50}). There is, however, an elegant way of generalising the results. We make use of the treatment of 
measurement theory sketched in Section \ref{bohm}, according to which after a measurement interaction one can identify a specific component of the wavefunction (\ref{8}) of the total 
system as uniquely responsible for the further motion of the particles. A measurement interaction, however, just means an appropriate choice of potentials and magnetic fields over a 
period of time, which we are free to make. 

Specifically, we first confine the particles via symmetric potentials to $n$ boxes with infinite walls, each box containing a single particle. We then perform (ideal) measurements of spin on 
each of the particles, yielding results, say, $s_1^0,\ldots,s_n^0$. The effective wavefunction for the $n$ particles is now
  \begin{equation}
    \sum_{\Pi}\psi_{s_{\Pi(1)}^0,\ldots,s_{\Pi(n)}^0} |s_{\Pi(1)}^0>\ldots |s_{\Pi(n)}^0>,
    \label{51}
  \end{equation}
where the sum ranges over all and only permutations that yield distinct sequences. Since with a fixed experimental arrangement, the different sequences of results depend on
different initial positions of the particles, the requirement that the original wavefunction leave all possible trajectories invariant under exchange of any two particles translates into 
requiring the same {\em separately} of all non-zero components of the form (\ref{51}). (If one such component is zero, then it is already totally symmetric or 
antisymmetric.)

For illustration, take the case of three spin-$1/2$ particles. The general case is analogous. (We need at least three particles in order to cover the possibility of parastatistics.) The easiest case is when all spin 
measurements yield the same result, say up. Then the effective wavefunction is
  \begin{equation}
    \psi_{+++}|+>|+>|+>.
      \label{52}
  \end{equation}
If we choose the magnetic field ${\bf B}=0$, the system is effectively described by one scalar wavefunction obeying the Schr\"{o}dinger
equation. We can open the boxes and treat it by the methods used for scalar wavefunctions in Section \ref{main}. 

In general, however, we will have different spin values in different boxes,
say spin up in two boxes and spin down in the third. We first pull down the potential barriers between boxes containing particles with the same spin, so that in this case we have one box 
containing two spin-up particles and one containing one spin-down particle. The effective wavefunction is now given by 
  \begin{equation}
    \psi_{++-}|+>|+>|-> + \psi_{+-+}|+>|->|+> + \psi_{-++}|->|+>|+>,
      \label{53}
  \end{equation}
each term corresponding to a different particle being in the spin-down box, and one still needs the 
guidance equations (\ref{48}) or (\ref{49}). Notice, however, that {\em as long as} the particles are confined to the two boxes, for any possible configuration only one of the terms 
in (\ref{53}) will be non-zero, e.g. for any configuration in which particle 1 is in the spin-down box, the only non-zero term is $\psi_{-++}|->|+>|+>$. It follows that, as long as the particles 
are in the boxes,
  \begin{eqnarray}
    \bfnabla S_{++-}({\bf x},{\bf y},{\bf z})    & = &   \bfnabla S_{++-}({\bf y},{\bf x},{\bf z}),   \label{54}  \\  
    \bfnabla S_{+-+}({\bf x},{\bf y},{\bf z})    & = &   \bfnabla S_{-++}({\bf y},{\bf x},{\bf z})    \label{55},    
  \end{eqnarray}
etc.\ (where $S_{++-}$ is the phase of $\psi_{++-}$, etc.). Taking ${\bf B}=0$, not only do $\psi_{++-}({\bf x},{\bf y},{\bf z})$ and 
$\psi_{++-}({\bf y},{\bf x},{\bf z})$ obey the same (symmetric) Schr\"{o}dinger equation, but also $\psi_{+-+}({\bf x},{\bf y},{\bf z})$ and 
$\psi_{-++}({\bf y},{\bf x},{\bf z})$, etc.

For each fixed $z$, when non-empty, the region $\{R_{++-}({\bf x},{\bf y},{\bf z})\neq 0\}$ is connected (we exchange the two particles in the spin-up box), so that we can apply our 
previous methods (for both phase and amplitude) to obtain
  \begin{equation}
    \psi_{++-}({\bf x},{\bf y},{\bf z}) = \pm\psi_{++-}({\bf y},{\bf x},{\bf z})
      \label{56}
  \end{equation}
(and similarly for $\psi_{+-+}$ under exchange of ${\bf x}$ and ${\bf z}$, and for $\psi_{-++}$ under exchange of ${\bf y}$ and ${\bf z}$). Instead, when 
non-empty, the region $\{R_{+-+}({\bf x},{\bf y},{\bf z})\neq 0\}\cup\{R_{-++}({\bf x},{\bf y},{\bf z})\neq 0\}$ has two disconnected components, according to whether particle 1 
or particle 2 is in the spin-down box. We obtain only
  \begin{equation}
    \psi_{+-+}({\bf x},{\bf y},{\bf z}) = \delta e^{i\gamma}\psi_{-++}({\bf y},{\bf x},{\bf z}),
      \label{57}
  \end{equation}
with $\gamma$ and $\delta$ arbitrary (and similarly in the other cases). One sees that the spinorial wavefunction must have the form
  \begin{eqnarray}
    \Psi ({\bf x},{\bf y},{\bf z})  & = &  \psi_{++-}({\bf x},{\bf y},{\bf z})|+>|+>|-> +                                         \nonumber   \\
                                                      &    &   \delta e^{i\gamma}\psi_{+-+}({\bf x},{\bf z},{\bf y})|+>|->|+> +      \nonumber   \\
                                                      &    &   \delta' e^{i\gamma'}\psi_{-++}({\bf z},{\bf y},{\bf x})|->|+>|+>,
     \label{58}
  \end{eqnarray}
with $\psi_{++-}$ symmetric or antisymmetric in the first two variables. 

As a penultimate step, we flip the spin in the spin-down box, set the magnetic field back to $0$, and pull down the wall 
between this box and the spin-up box. (If there are more than two boxes with different values of spin, we repeat this operation several times.) The wavefunction in (\ref{58}) becomes 
  \begin{equation}
    \Big(\psi_{++-}({\bf x},{\bf y},{\bf z}) + \delta e^{i\gamma}\psi_{+-+}({\bf x},{\bf z},{\bf y}) + \delta' e^{i\gamma'}\psi_{-++}({\bf z},{\bf y},{\bf x})\Big)|+>|+>|+>,
     \label{59}
  \end{equation}
i.e. has the form (\ref{52}), and must therefore be symmetric in all pairs of variables (if $\psi_{++-}$ is symmetric in the first two variables) or antisymmetric in all pairs of variables
(if $\psi_{++-}$ is antisymmetric in the first two variables). It follows easily that, respectively, 
  \begin{equation}
     \delta e^{i\gamma} = \delta' e^{i\gamma'} = \pm 1.
     \label{60}
  \end{equation}
Thus, we have shown that also an effective wavefunction 
with different spin terms such as (\ref{53}) is totally symmetric or antisymmetric. 

Finally, one convinces oneself that the effective wavefunctions need to be either all symmetric or all 
antisymmetric (e.g.\ by considering non-maximal measurements after which the effective wavefunction is a superposition of any two thus far considered). Thus, the spinorial wavefunction 
is totally symmetric or totally antisymmetric at the time of the spin measurements, and since the Pauli equation is symmetric, for all times. 

This completes our derivation of the symmetry conditions in the spinorial case, and shows that parastatistics is excluded. Unfortunately, our proof yields no complete connection between
spin and statistics, since for arbitrary spin the velocity field is invariant both when the wavefunction is symmetric and when it is antisymmetric. Incidentally, there is another 
model of spin in pilot-wave theories, due to Bohm and Hiley 
(1993, Chap.\ 10), where the guidance equation includes an additional velocity term. (This term is added so to have 
the velocity field obtained from the Pauli equation be the non-relativistc limit of the one obtained from the Dirac equation.) However, this term also transforms invariantly under exchange 
irrespectively of whether the wavefunction is symmetric or antisymmetric, so that Bohm and Hiley's model can also not be used to derive the spin-statistics connection.

\section{Alternative approaches}\label{topological}
To conclude, we briefly mention the alternative approaches to identical particles in non-relativistic quantum mechanics, one of which also aims to exclude parastatistics and bears 
an interesting relation to our approach above. This is the {\em topological approach} developed by Laidlaw and C. DeWitt (1971)
and by Leinaas and Myrheim (1977), 
with an anticipation by Finkelstein and Rubinstein (1968). 
This approach is based on the use of a {\em reduced configuration space} for identical particles, defined by excising the points in which two or more particles coincide, and identifying 
configurations related by a permutation of the particles. Exclusion of the coincidence points makes the resulting space multiply connected. In three dimensions or more, it is doubly 
connected, corresponding to the simply connected case above (without identifications). In two dimensions (treated only by Leinaas and Myrheim), the reduced configuration space is 
infinitely connected. 

Laidlaw and DeWitt use the Feynman-path formalism to calculate transition amplitudes. From the topology of the paths, they derive the 
usual amplitudes. (Notice the different point of application of the topological considerations: Feynman paths versus equal-time paths.) Leinaas and Myrheim  
define wavefunctions (and a Schr\"{o}dinger equation) directly on the reduced configuration space, and obtain two-valued and multi-valued wavefunctions in three and two 
dimensions, respectively (they also have a discussion of the one-dimensional case). Laidlaw and DeWitt treat only the spinless case, while Leinaas and Myrheim give an extension 
to spinor-valued wavefunctions on the reduced configuration space (explicitly only for two particles).

In both cases, the choice of the reduced configuration space is conceptually the central step. As pointed out already in Brown {\em et al.\ }(1999),
the pilot-wave approach gives a 
particularly perspicuous justification for this step. Indeed, the set of coincidences is dynamically inaccessible, whether or not the wavefunction is zero on the set. This can be seen 
as follows. We set $\xi_1=\xi_2={\bf x}$ in (\ref{12}) above:
  \begin{equation}
    \bfnabla_1 S(\xi_1,\xi_2,t)+{\bf A}(\xi_1,t)\Big|_{\xi_1=\xi_2={\bf x}} = \bfnabla_2 S(\xi_1,\xi_2,t)+{\bf A}(\xi_2,t)\Big|_{\xi_1=\xi_2={\bf x}},
    \label{61}
  \end{equation}
and see that if two particles coincide at one instant $t$, they coincide for all times. Thus, if they do not coincide at one instant, they cannot come to coincide at any later time.
The configuration space for two identical particles decomposes into an unphysical configuration space for one particle of double mass and charge, and the configuration space 
with coincidence points removed. As regards the identification of points related by permutations, it would seem that this is also perfectly natural, since the motions are perfectly 
symmetrical, and indeed de Broglie--Bohm theory can be consistently formulated in a label-free way, as described in D\"{u}rr {\em et al.\ }(in preparation)
(see also Sj\"{o}qvist and Carlsen, 1995).

Goldin {\em et al.\ }(1980, 1981)
work within the framework of second-quantised non-relativistic fields $\psi({\bf x})$, or more 
precisely, the Lie algebra formed by the mass density and momentum density operators: 
  \begin{equation}
    \begin{array}{lcl}
      \rho({\bf x})       & := &   m \psi^*({\bf x})\psi({\bf x}),   \\[1em]
      {\bf J}({\bf x})    & := &   
      \frac{\hbar}{2i}(\psi^*({\bf x})\bfnabla\psi({\bf x})-\bfnabla\psi^*({\bf x})\psi({\bf x}))
    \end{array}
    \label{62}
  \end{equation}
(suitably averaged with Schwartz space functions). The representations of this {\em current algebra} yield fields satisfying the canonical commutation and anticommutation relations, 
as well as anyonic commutation relations (see also Goldin and Sharp, 1983,
where the relation to Leinaas and Myrheim's work
is discussed in more detail). Goldin (1987) points out that also this approach yields a natural justification for excising the coincidence points. 
As opposed to the justification provided by de Broglie--Bohm theory, however, coincidence points are
excised also in the case of distinguishable particles. 

The current algebra approach provides a rigorous framework for anyons, but it explicitly allows for parastatistics, unless the wavefunction is scalar-valued (Goldin, 1987). 
Goldin {\em et al.\ }(1985) therefore conjecture that the topological approach may be able to rigorously exclude parastatistics only in the case of scalar wavefunctions (e.g.\ suitable
extensions of Feynman-path methods would also allow for parastatistics). If this is correct, the approach based on pilot-wave theories, as in the present paper and in D\"{u}rr
{\em et al.\ }(in preparation) would provide the first rigorous proof that parastatistics are excluded also in the spinorial case.

\section*{Acknowledgements}
I wish to thank first of all Harvey Brown and Erik Sj\"{o}qvist for early discussions of this topic in the context of our joint paper (Brown {\em et al.}, 1999).
I am further very much indebted to Detlef D\"{u}rr, Shelly Goldstein, James Taylor and Nino Zangh\`{\i} for discussion, correspondence and earlier drafts of D\"{u}rr {\em et al.\ }(in preparation); 
to Detlef also for attracting my attention to Schneider (1995)
and for a week's wonderful hospitality at the University of Munich, and to Shelly and
especially James for pointing out an error in a previous draft. I am grateful to Simon Saunders and Antony Valentini for discussion, 
and to Jerry Goldin for yet more discussion, correspondence and various useful references and papers. I would 
finally like to thank audiences at the Seventh U.K.\ Conference on Foundations of Physics, the Sixth Annual Florence--Stanford Meeting, the 2001 Biennial Meeting of the International Quantum Structures Association, and at the Universities of Oxford, Utrecht, Freiburg, M\"{u}nchen and at the 
Technische Universit\"{a}t M\"{u}nchen. I completed part of this work as holder of a generous Postdoctoral Research Fellowship 
from the British Academy, and part as the spoilt and happy guest of Carsten Held in the Philosophisches Seminar I at the University of Freiburg.

\section*{References}

\noindent Arovas, D., Schrieffer, J.\ R., and Wilczek, F.\ (1982), `Fractional Statistics and
the Quantum Hall Effect', {\em Physical Review Letters} 
{\bf 53}, \mbox{722--723}.
 
\noindent Bell, J.\ S.\ (1987), {\em Speakable and Unspeakable in Quantum Mechanics} (Cambridge:
Cambridge University Press).

\noindent Bohm, D.\ (1952), `A Suggested Interpretation of the Quantum Theory in Terms of 
``Hidden'' Variables. I' and `II', {\em Physical Review} {\bf 85}, \mbox{166--179} and 
\mbox{180--193}.

\noindent Bohm, D., and Hiley, B.\ J.\ (1993), {\em The Undivided Universe} (London: Routledge).


\noindent Broglie, L.\ de (1928), 'La nouvelle dynamique des quanta', in {\em \'{E}lectrons et 
Photons, Rapports et discussions du cinqui\`{e}me Conseil de Physique Solvay} (Paris, 
Gauthier-Villars), \mbox{pp. 105--141}. 

\noindent Brown, H.\ R., Sj\"{o}qvist, E., and Bacciagaluppi, G.\ (1999), `Remarks on Identical 
Particles in de Broglie--Bohm Theory', {\em Physics Letters}  {\bf A 251}, \mbox{229--235}.

\noindent D\"{u}rr, D.\ (2001), {\em Bohmsche Mechanik als Grundlage der Quantenmechanik} (Berlin:
Springer).

\noindent D\"{u}rr, D., Goldstein, S., Taylor, J., and Zangh\`{\i}, N. (in preparation), `Bosons, Fermions,
and the Natural Configuration Space of Identical Particles'.

\noindent Finkelstein, D., and Rubinstein, J.\ (1968), `Connection between Spin, Statistics,
and Kinks', {\em Journal of Mathematical Physics} {\bf 9}, \mbox{1762--1779}.

\noindent Goldin, G.\ A.\ (1987), `Parastatistics, $\theta$-statistics, and Topological Quantum 
Mechanics from Unitary Representations of Diffeomorphism Groups', in H.-D.\ Doebner and J.\ D.\ 
Hennig (eds), {\em Proceedings of the XV International Conference on Differential Geometric 
Methods in Theoretical Physics} (Singapore: World Scientific), \mbox{pp. 197--207}.
    
\noindent Goldin, G.\ A., Menikoff, R., and Sharp, D.\ H.\ (1980), `Particle Statistics from
Induced Representations of a Local Current Group', {\em Journal of Mathematical 
Physics} {\bf 21}, \mbox{650--664}.

\noindent Goldin, G.\ A., Menikoff, R., and Sharp, D.\ H.\ (1981), `Representations of a Local 
Current Algebra in Nonsimply Connected Space and the Aharonov--Bohm Effect', {\em Journal of Mathematical 
Physics} {\bf 22}, \mbox{1664--1668}.

\noindent Goldin, G.\ A., Menikoff, R., and Sharp, D.\ H.\ (1985), `Comments on ``General Theory
for Quantum Statistics in Two Dimensions'' ', {\em Physical Review 
Letters} {\bf 54}, \mbox{603}.

\noindent Goldin, G.\ A., and Sharp, D.\ H.\ (1983), `Rotation Generators in Two-Dimensional
Space and Particles Obeying Unusual Statistics', {\em Physical Review} {\bf D 28}, \mbox{830--832}.

\noindent Goldstein, S.\ (1987), `Stochastic Mechanics and Quantum Theory', {\em Journal of 
Statistical Physics} {\bf 47}, \mbox{645--667}.

\noindent Goldstein, S.\ (2001), `Bohmian Mechanics', in {\em The Stanford Encyclopedia of 
Philosophy}, http://plato.stanford.edu/entries/qm-bohm/ .

\noindent Holland, P.\ R.\ (1993), {\em The Quantum Theory of Motion} (Cambridge: Cambridge 
University Press).

\noindent Laidlaw, M.\ G.\ G., and DeWitt, C.\ M.\ (1971), `Feynman Functional Integrals for
Systems of Indistinguishable Particles', {\em Physical Review} {\bf D 3}, \mbox{1375--1378}. 

\noindent Laughlin, R.\ B.\ (1983), `Anomalous Quantum Hall Effect: An Incompressible Quantum
Fluid with Fractionally Charged Excitations', {\em Physical Review Letters} {\bf 50}, 
\mbox{1395--1402}.
 
\noindent Leinaas, J.\ M., and Myrheim, J.\ (1977), `On the Theory of Identical Particles',
{\em Il Nuovo Cimento} {\bf B 37}, \mbox{1--23}.
  
\noindent Messiah, A.\ M.\ L., and Greenberg, O.\ W.\ (1964), `Symmetrization Postulate and
its Experimental Foundation', {\em Physical Review} {\bf 136 B}, \mbox{248--267}. 

\noindent Nelson, E.\ (1966), `Derivation of the Schr\"{o}dinger Equation from Newtonian
Mechanics', {\em Physical Review} {\bf 150}, \mbox{1079--1085}.

\noindent Nelson, E.\ (1985), {\em Quantum Fluctuations} (Princeton: Princeton University Press).

\noindent Schneider, M.\ (1995), {\em Identische Teilchen und Bohmsche Mechanik}, Diplomarbeit, 
Ludwig-Maximilians-Universit\"{a}t M\"{u}nchen, Sektion Physik,\\ 
http://emma.informatik.unibw-muenchen.de/~schneider/ .

\noindent Sj\"{o}qvist, E., and Carlsen, H.\ (1995), `Fractional Statistics in Bohm's Theory',
{\em Physics Letters} {\bf A 202}, \mbox{160--166}.
 
\noindent Tsui, D.\ C., Stormer, H.\ L., and Gossard, A.\ C.\ (1982), `Two-Dimensional
Magnetotransport in the Extreme Quantum Limit', {\em Physical Review 
Letters} {\bf 48}, \mbox{1559--1562}. 

\noindent Valentini, A.\ (in preparation), {\em Pilot-Wave Theory} (Cambridge: Cambridge 
University Press).

\noindent Wilczek, F.\ (1982), `Quantum Mechanics of Fractional-Spin Particles', {\em Physical 
Review Letters} {\bf 49}, \mbox{957--959}.
 

\end{document}